\documentclass[a4paper]{article}
\usepackage[psamsfonts]{amssymb}
\usepackage{amsmath}
\usepackage{pstricks}
\usepackage{cite}
\date{}

\author{M. Alimohammadi\footnote{alimohmd@ut.ac.ir}\ \ and L. Sadeghian\footnote{lsadeghian@ut.ac.ir}
\\ {\small Department of Physics, University of Tehran,}
\\ {\small North Karegar Ave., Tehran, Iran.}}
\title{ Quantum induced $\omega =-1$ crossing of the quintessence and phantom models}
\begin{document}
\maketitle
\begin{abstract}

Considering the single scalar field models of dark energy, i.e.
the quintessence and phantom models, it is shown that the quantum
effects can cause the system crosses the $\omega=-1$ line. This
phenomenon does not occur in classical level. The quantum effects
are described via the account of conformal anomaly.
\end{abstract}

\section{Introduction}
One of the most important aspects of the present universe is its
accelerated expansion. Since two independent observations based on
redshift-distance relation of type Ia supernovas in 1998
\cite{Riess,Perlmutter}, numerous observations \cite{quintum1}
consistently indicate that our universe is dominated by a perfect
fluid with negative pressure, dubbed dark energy, which
constitutes two third of the present universe.

The first candidate which has been introduced for dark energy is a
cosmological constant $\Lambda$ of order $(10^{-3}\ {\rm eV} )^4$,
with equation of state parameter $\omega={p}/{\rho}=-1$. This
model suffers from fine tuning and coincidence problems
\cite{Weinberg}. As an alternative to cosmological constant, the
dynamical models have been introduced. In these models, the
equation of state parameters $\omega(z)$ is one of the main
parameters which is usually used in studying the time variation of
dark energy. In accelerating universe, $\omega$ satisfies $\omega<
-1/3$.

In the quintessence model of dark energy, which consists of one
normal scalar field $\varphi$ \cite{Asym6}, $\omega$ is always
$\omega>-1$. In phantom model, which is a scalar field theory with
a field $\sigma$ with negative kinetic energy, $\omega$ always
satisfies $\omega<-1$ \cite{Asym7}. But some astrophysical data
seem to slightly favor an evolving dark energy and show a recent
$\omega=-1$, the so-called phantom-divide-line, crossing
\cite{Asym5}. This phenomenon can not be explained by none of
these two models, the quintessence or phantom models. A possible
way to overcome this problem is to consider two scalar fields in
the models known as hybrid models. One of these models is the
model consists of one quintessence and one phantom field, the
so-called quintom model \cite{Asym8}. Recently, it has been shown
that the $\omega>-1$ to $\omega<-1$ transition always occurs in
the quintom models with slowly-varying potentials \cite{Asym9}.
Also if one considers one scalar field, but with suitable
interaction with background dark matter, again this transition can
be occurred \cite{Asym10}.

In present paper, we study the contribution of quantum effects in
$\omega=-1$ crossing of single scalar field models of dark energy,
that is the quintessence and phantom models. Due to a No-Go
theorem proposed in \cite{nogo}, a single scalar field which
minimally couples to Einstein gravity can cross the
phantom-divide-line only when the higher derivative terms of
scalar field, like $\varphi \square\varphi$ \cite{highder}, exist
in the Lagrangian. But, as we see, it is not the case at the
quantum level. This transition can be induced
quantum-mechanically, with no need to higher derivative terms. The
quantum effects are described via the account of conformal
anomaly, reminding about anomaly-driven inflation \cite{Star}. The
contribution of this quantum effect in preserving the most of the
energy conditions of phantom matter has been discussed in
\cite{Nojiri1} and its influence in moderating the sudden future
singularity (Big Rip) of phantom model has been studied in
\cite{Nojiri2}.

The scheme of the paper is as follows. In section 2 we briefly
review the quintessence and phantom models and introduce the
perturbative method of studying the phantom-divide-line crossing
of these models. The energy density and pressure resulting from
the conformal anomaly are also quoted. In section 3 we apply our
method to quintessence and phantom models and show that the
system, except for very special initial conditions, has a
transition from $\omega<-1$ to $\omega>-1$, or vise versa,
resulting from quantum effects. In special free pure phantom
model, it is shown that this transition always occurs from
$\omega<-1$ region to $\omega>-1$ region.

We use the units $\hbar=c=G=1$ throughout the paper.

\section{ Perturbative method for studying the $w=-1$ crossing}

Consider a spatially flat Friedman-Robertson-Walker space-time in
co-moving coordinates $(t,x,y,z)$
\begin{equation} \label{1}
\textrm{d}s^{2}=-\textrm{d}
t^{2}+a^{2}(t)(\textrm{d}x^{2}+\textrm{d}y^{2}+\textrm{d}z^{2})
\end{equation}
where $a(t)$ is the scale factor. It is assumed that the universe
is filled with (dark) matter and a single scalar field. The
evolution equation of matter density $\rho_m$ is
\begin{equation}\label{2}
\dot{\rho}_m+3H\gamma_m\rho_m=0,
\end{equation}
in which $ \gamma_m=1+\omega_m$. $\omega_m$, the equation of state
parameter of matter field, is defined through
$\omega_m={p_m}/{\rho_m}$, in which $p_m$ is the pressure of
matter field. $H(t)={\dot{a}(t)}/{a(t)}$ is the Hubble parameter
and "dot" denotes the time derivative. The dark energy consists of
the quintessence field $\varphi$ (or phantom field $\sigma$),
which in the case of homogenous field, its energy density $\rho_D$
and pressure $p_D$ ("$D$" denotes the dark energy) are
\cite{Asym6,Asym7}
\begin{equation}\label{3}
\rho_{\mathrm{quintessence}}=\frac{1}{2}\dot{\varphi}^2+V(\varphi)\,
\,\, ,\, \, \,
p_{\mathrm{quintessence}}=\frac{1}{2}\dot{\varphi}^2-V(\varphi),
\end{equation}
or
\begin{equation}\label{4}
\rho_{\mathrm{phantom}}=-\frac{1}{2}\dot{\sigma}^2+V(\sigma)\,\,\,
,\,\,\, p_{\mathrm{phantom}}=-\frac{1}{2}\dot{\sigma}^2-V(\sigma).
\end{equation}
The Friedman equations are
\begin{equation}\label{5}
H^2=\frac{8\pi}{3}\rho_{\textrm{tot.}},
\end{equation}
and
\begin{equation}\label{6}
\dot{H}=-4\pi(\rho_{\textrm{tot.}}+p_{\textrm{tot.}}).
\end{equation}
The equation of state parameter
$\omega={p_{\mathrm{tot.}}}/{\rho_{\mathrm{tot.}}}$ is found as
\begin{equation}\label{7}
\omega=-1-\frac{2}{3}\frac{\dot{H}}{H^2}.
\end{equation}
For quintessence phase $\omega>-1$, we have $\dot{H}<0$ and in
phantom phase $\omega<-1$, $\dot{H}$ obeys $\dot{H}>0$. If
$\dot{H}(t_0)=0$ and $H(t)$ has a relative extremum at $t=t_0$,
the system crosses $\omega=-1$ line at time $t=t_0$

If we restrict ourselves to $t-t_0\ll h_0^{-1}$, where
$h_0=H(t_0)$ and $h_0^{-1}$ is of order of the age of universe,
$H(t)$ can be expanded as
\begin{equation}\label{8}
H(t)=h_0+h_1(t-t_0)^{\alpha}+h_2(t-t_0)^{\alpha+1}+O\left((t-t_0)^{\alpha+2}\right).
\end{equation}
$\alpha\geq 2$ is the order of first non-vanishing derivative of
$H(t)$ at $t=t_0$ and $h_1=\frac{1}{\alpha !}H^{(\alpha)}(t_0)$.
$H^{(n)}(t_0)$ is the $n$-th derivative of $H(t)$ at $t=t_0$. The
transition from $\omega>-1$ region to $\omega<-1$ region occurs
when $\alpha$ is even positive integer and $h_1>0$. In reverse
case, $h_1$ must be negative. In the case of quintom model, it has
been shown that for slowly-varying potentials $V(\varphi,\sigma)$,
$\alpha=2$ and $h_1>0$, and therefore $\omega>-1$ to $\omega<-1$
transition occurs \cite{Asym9}.

To consider the quantum effects, one may use a standard method
which leads to a closed form for quantum corrections. In this
method, the interaction is considered between the quantum free
matter field and classical gravitational field
\cite{Birrell,Star}. It can be seen that the renormalization of
effective action leads to some extra terms in the trace of
energy-momentum tensor, which is known as trace/conformal anomaly.
Note that in the classical level, the energy-momentum tensor is
traceless. These extra terms are given by:
\begin{eqnarray} \label{81}
T=b(F+\frac{2}{3}\Box R)+b'G+b''\Box R,
\end{eqnarray}
where $F$ is the square of 4d Weyl tensor and $G$ is Gauss-Bonnet
invariant, given by:
\begin{eqnarray} \label{82}
&&F=\frac{1}{3}R^2-2R_{ij}R^{ij}+R_{ijkl}R^{ijkl}, \nonumber\\
&&G=R^2-4R_{ij}R^{ij}+R_{ijkl}R^{ijkl}.
\end{eqnarray}
Generally for $N$ scalars, $N_{1/2}$ spinors, $N_1$ vector fields,
$N_2(=0 \:\: \textrm{or}\:\: 1)$ gravitons and $N_{HD}$ higher
derivative conformal scalars (including phantom), $b$, $b'$ and
$b''$ are given by
\begin{eqnarray} \label{11}
&&b=\frac{N+6N_{1/2}+12N_1+611N_2-8N_{HD}}{120(4\pi)^2},\nonumber\\
&&b'=-\frac{N+11N_{1/2}+62N_1+1411N_2-28N_{HD}}{360(4\pi)^2}\,\,
,\,\, b''=0.
\end{eqnarray}
Using eq.(\ref{81}), one can find the contributions due to
conformal anomaly to $\rho$ and $p$ as follows \cite{Nojiri3}
\begin{eqnarray} \label{9}
\rho_A&=&-\frac{1}{a^4}\left\{ b'(6a^4H^4+12a^2H^2)\right.\nonumber \\
&&+(\frac{2}{3}b+b'')\left[a^4(-6H{\ddot
{H}}-18H^2{\dot{H}}+3{\dot{H}}^2) +6a^2H^2\right]\nonumber \\
&&\left.-2b+6b'-3b''\right\},
\end{eqnarray}
and
\begin{eqnarray} \label{10}
p_A&=&b'\left[
6H^4+8H^2{\dot{H}}+\frac{1}{a^2}(4H^2+8{\dot{H}})\right]\nonumber \\
&&+\left( \frac{2}{3}b+b''\right)[-2{\dddot{H}}-12H{\ddot{H}}
-18H^2{\dot{H}}-9{\dot{H}}^2\nonumber\\
&&+\frac{1}{a^2}(2H^2+4{\dot{H}})]-\frac{-2b+6b'-3b''}{3a^4}.
\end{eqnarray}
Now it looks reasonable to solve the Friedman equations with these
quantum corrections and see if there exists any new result in the
phantom-divide-line-crossing issue due to this correction.

\section{The transition solutions}

In this section we consider the expansion (\ref{8}) for $H(t)$ and
try to solve the equations (\ref{2}), (\ref{5}) and (\ref{6}) with
$\rho_{\textrm{tot.}}=\rho_m+\rho_D+\rho_A$ and
$p_{\textrm{tot.}}=p_m+p_D+p_A$. We want to find any consistent
solution of these equations with $\omega=-1$ crossing property.

\subsection{The quintessence model}

In the case of quintessence field, one has $N=1$ and
$N_{1/2}=N_1=N_2=N_{HD}=0$. So
\begin{equation}\label{12}
b=-3b'=\frac{1}{120{(4\pi)^2}}.
\end{equation}
Eqs. (\ref{2}) and (\ref{8}) (with $\alpha\geq 2$ and $t_0\equiv
0$) result in
\begin{equation}\label{13}
\rho_m(t)=\rho_m(0)[1-3h_0\gamma_mt+\frac{9}{2}\gamma_m^2h_0^2t^2+...].
\end{equation}
By expanding both sides of eq.(\ref{5}) near $t_0=0$, one finds
\begin{equation}\label{14}
h_0^2+2h_0h_1t^\alpha+...=\frac{8\pi}{3}[\rho_{\textrm{tot.}}(0)+\dot\rho_{\textrm{tot.}}(0)t
+\frac{1}{2}\ddot\rho_{\textrm{tot.}}t^2+...],
\end{equation}
in which
\begin{equation}\label{15}
\rho_{\textrm{tot.}}(t)=\rho_{\textrm{quintessence}}+\rho_m+\rho_A=\rho_{\textrm{cl.}}+\rho_A.
\end{equation}
In above equation, "cl." denotes "classical". Eq.(\ref{14}) then
results in the following two relations:
\begin{equation}\label{16}
h_0^2=\frac{8\pi}{3}\rho_{\textrm{tot.}}(0)= \frac{8\pi}{3} \left[
\rho_{\textrm{cl.}}(0)+2b\left( h_0^4+4h_0h_1\delta_{\alpha
,2}+\frac{2}{a_0^4}\right)\right],
\end{equation}
\begin{equation}\label{17}
0=\dot{\rho}_{\textrm{tot.}}(0)=\dot{\rho}_{\textrm{cl.}}(0) + 2b
\left( 12h_0h_2\delta_{\alpha ,2}+12h_0^2h_1\delta_{\alpha
,2}+12h_0h_1\delta_{\alpha ,3}-\frac{8h_0}{a_0^4}\right),
\end{equation}
in which $a_0$ is the scale factor at transition time $t=0$. The
same expansion for the second Friedman equation (\ref{6}) results
in the following extra relation:
\begin{equation}\label{19}
0=\delta(0)=
\rho_{\textrm{cl.}}(0)+p_{\textrm{cl.}}(0)-\frac{4b}{3}\left(
6h_0h_1\delta_{\alpha ,2}+6h_2\delta_{\alpha
,2}+6h_1\delta_{\alpha ,3}-\frac{4}{a_0^4}\right),
\end{equation}
in which
\begin{equation}\label{21}
\delta(t)=\rho_{\textrm{tot.}}+p_{\textrm{tot.}}=\rho_{\textrm{cl.}}+p_{\textrm{cl.}}+\rho_{A}+p_{A}.
\end{equation}
In eqs.(\ref{16})-(\ref{19}), $\rho_m(t)$ is given by
eq.(\ref{13}) and $\rho_m+p_m=\gamma_m\rho_m$.
$\rho_{\textrm{quintessence}}$ and $p_{\textrm{quintessence}}$ are
those in eq.(\ref{3}), $\rho_A$ and $p_A$ are given by
eqs.(\ref{9}) and (\ref{10}) with $H(t)=h_0+h_1t^\alpha+...$, $b$
and $b'$ from eq.(\ref{12}), and $b''=0$. Note that eqs.(\ref{17})
and (\ref{19}) indicate the energy conservation law
$\dot{\rho}_{\textrm{tot.}}+3H(\rho_{\textrm{tot.}}+p_{\textrm{tot.}})=0$
at $t=0$. Since $\delta
(0)=\rho_{\textrm{tot.}}(0)+p_{\textrm{tot.}}(0)=0$,
$\dot{\rho}_{\textrm{tot.}}$ must satisfy
$\dot{\rho}_{\textrm{tot.}}(0)=0$.

Let us first consider the equation (\ref{19}). Since
$\rho_{\textrm{cl.}}+p_{\textrm{cl.}}=\dot{\varphi}^2+\gamma_m\rho_m$,
eq.(\ref{19}) for $\alpha\geq 4$ results in:
\begin{equation}\label{n22}
\dot{\varphi}^2(0)+\gamma_m\rho_m(0)+\frac{16b}{3a_0^4}=0 \,\,\, ,
\,\,\, (\textrm{for}\,\, \alpha\geq 4).
\end{equation}
The above equation has no solution except $\dot{\varphi}(0)=0$,
$\rho_m(0)=0$, and $a_0\rightarrow\infty$, which is unphysical. So
we have only two choices $\alpha=2$ and $\alpha=3$.

For $\alpha=2$, three equations (\ref{16})-(\ref{19}) can be used
to obtain the coefficients $h_0$, $h_1$ and $h_2$. $h_0$ is found
as
\begin{equation}\label{n23}
h_0=-\frac{\ddot{\varphi}(0)+(\textrm{d} V/\textrm{d}
\varphi)_0}{3\dot\varphi(0)},
\end{equation}
which is nothing but the evolution equation
\begin{equation}\label{n24}
\ddot{\varphi}+3H\dot{\varphi}+\frac{\textrm{d}
V(\varphi)}{\textrm{d} \varphi}=0
\end{equation}
at $t=0$. $h_1$, in terms of $h_0$, is found from eq.(\ref{16}) as
following \
\begin{equation}\label{n25}
h_1=\frac{1}{8bh_0}\left\{\frac{3}{8\pi}h_0^2-\left[\rho_{\textrm{cl.}}
(0)+2bh_0^4+\frac{4b}{a_0^4}\right]\right\},
\end{equation}
in which
$\rho_{\textrm{cl}.}(0)=\frac{1}{2}\dot\varphi^2(0)+V(0)+\rho_m(0)$.
Finally, $h_2$ can be expressed in terms of $h_1$ by using
eqs.(\ref{17}) or (\ref{19}). The next-leading relations from
Friedman equations, that is the coefficients of $t^2$ of
eq.(\ref{5}) and $t$ of eq.(\ref{6}), are:
\begin{equation}\label{n26}
2h_0h_1=\frac{1}{2}\ddot{\rho}_{\textrm{tot.}}(0).
\end{equation}
and
\begin{equation}\label{n27}
2h_1=-4\pi\dot{\delta}(0).
\end{equation}
Each of the above equations can be used to determine the parameter
$h_3$ of expansion (\ref{8}), which are the same if one uses the
field equation (\ref{n24}).

It is interesting to note that the quantum correction terms (the
third and fourth terms in the right-hand-side of eq.(\ref{n25}))
are much smaller than the classical terms:
\begin{equation}\label{25}
h_0^4\sim\frac{1}{a_0^4}\ll{h_0^2}.
\end{equation}

The important observation is that if we set $b\rightarrow0$, then
eq.(\ref{19}) results in $\dot\varphi(0)=0$ and $\rho_m(0)=0$,
from which eq.(\ref{n27}) reduces to $h_1=0$ (note that
$\delta(0)=\dot\varphi^2(0)+\gamma_m\rho_m(0)$ and
$\dot\delta(0)=2\dot\varphi(0)\ddot\varphi(0)-3h_0\gamma_m^2\rho_m(0)$).
So without quantum effects, there is no non-trivial solution for
eqs.(\ref{5}) and (\ref{6}) with $H(t)$ given by eq.(\ref{8}), and
therefore there is no phantom-divide-line crossing in quintessence
model in classical level. But by considering the quantum effects,
$h_1$ has non-trivial solution (\ref{n25}) which can be positive
or negative, depending on the values $h_0$, $a_0$,
$\dot\varphi(0)$, $V(0)$ and $\rho_m(0)$. So quantum effects can
induce the $\omega=-1$ crossing in quintessence models.

As pointed out earlier, it has been shown that the behavior of
finite-time singularity in phantom models becomes rather milder if
one considers the quantum corrections. This is because near the
singularity, the curvatures and their time derivatives become
larger, so the quantum corrections, which include the powers of
these quantities, become large and important and can control the
singularity. So one can expect that the conformal anomaly, which
in our considered quintessence model is the only reason for
transition from quintessence phase to phantom phase, can not
itself cause any Big Rip.

Now consider the possibility of $\alpha=3$. In this case, one
again finds the field equation (\ref{n23}), which expresses the
Hubble parameter $H(t)$ at $t=0$ in terms of $\varphi (0)$,
$\dot{\varphi} (0)$ and $\ddot{\varphi} (0)$. But now the first
equation, eq.(\ref{16}), also gives $h_0$ in terms of the field
$\varphi$ and its derivatives, the matter density $\rho_m$, and
the scale factor $a(t)$ at $t=0$:
\begin{equation}\label{n29}
h_o=\left\{ \frac{1}{2}\left[ \frac{3}{16\pi b}\pm\sqrt{ \left(
\frac{3}{16\pi b}\right)^2-\frac{2}{b}\left( \rho_{\textrm{cl.}}
(0) +\frac{4b}{a_0^4}\right) }\right]\right\}^{1/2}.
\end{equation}
So $\alpha=3$ solution exists only if the right-hand-sides of
eqs.(\ref{n23}) and (\ref{n29}) are equal, which is a very special
choice of initial values. Under these conditions, of course, there
is no $\omega =-1$ transition. So except these fine-tuned initial
values, the solution is $\alpha=2$ and the quantum effects induce
the $\omega=-1$ crossing in quintessence models.

It is worth noting that this quantum phenomenon may have a
contribution equal to, or even more important than, the classical
effects if one considers the early stages of the universe in which
$h_0$ is large ( see, for example, eq.(\ref{n25})). This is, of
course, consistent with our physical intuition about the role of
quantum effects in gravitational phenomena.

\subsection{The phantom model}

To study the phantom model one must consider $N=N_{1/2}=N_1=N_2=0$
and $N_{HD}=1$ in eq.(\ref{11}). So
\begin{equation}\label{26}
b'=-\frac{7}{6}b=\frac{7}{90(4\pi)^2}.
\end{equation}
The procedure of previous subsection can be followed here. In this
way we find equations similar to (\ref{16})-(\ref{19}), but now
$\rho_D$ and $p_D$ are those introduced in eq.(\ref{4}) and
$\rho_A$ and $p _A$ are given by eqs.(\ref{9}) and (\ref{10}) with
$b'=-(7/6)b$ and $b''=0$. The result is:
\begin{equation}\label{n31}
h_0^2= \frac{8\pi}{3} \left[ \rho_{\textrm{cl.}}(0)-\frac{6b'}{7}
\left( 7h_0^4+\frac{10h_0^2}{a_0^2}+8h_0h_1\delta_{\alpha
,2}+\frac{9}{a_0^4}\right)\right],
\end{equation}
\begin{equation}\label{n32}
0=\dot{\rho}_{\textrm{cl.}}(0) -\frac{6b'}{7} \left(
24h_0h_2\delta_{\alpha ,2}+24h_0^2h_1\delta_{\alpha
,2}+24h_0h_1\delta_{\alpha
,3}-\frac{20h_0^3}{a_0^2}-\frac{36h_0}{a_0^4}\right),
\end{equation}
and
\begin{equation}\label{n33}
0=\rho_{\textrm{cl.}}(0)+p_{\textrm{cl.}}(0)+\frac{8b'}{7}\left(
6h_0h_1\delta_{\alpha ,2}+6h_2\delta_{\alpha
,2}+6h_1\delta_{\alpha
,3}-\frac{5h_0^2}{a_0^2}-\frac{9}{a_0^4}\right),
\end{equation}
in which
\begin{equation}\label{n34}
\rho_{\textrm{cl.}}= -\frac{1}{2}\dot{\sigma}^2 +V(\sigma) +\rho_m
\,\,\, , \,\,\,  \rho_{\textrm{cl.}}+p_{\textrm{cl.}} =
-\dot{\sigma}^2+\gamma_m\rho_m.
\end{equation}
For $\alpha\geq 3$, eq.(\ref{n31}) does not depend on $h_1$, so
$h_0$ is determined from two independent equations. The first one
is the equation of motion of phantom field
\begin{equation}\label{27}
h_0=-\frac{\ddot\sigma(0)-(\textrm{d}V/{\textrm{d}\sigma})_0}{3\dot\sigma(0)},
\end{equation}
and second one is eq.(\ref{n31}). These two expressions are equal
only if a very specific initial conditions has been chosen. So the
only typical solution, which always exists, is $\alpha=2$.

In the case $\alpha=2$, the equations (\ref{n31})-(\ref{n33}) can
be used to calculate $h_0$, $h_1$ and $h_2$. $h_0$ is specified by
eq.(\ref{27}), and $h_1$ and $h_2$ can be expressed in terms of
$h_0$. Using (\ref{n31}), $h_1$ is found as:
\begin{equation}\label{28}
h_1=\frac{7}{48b'h_0}\left\{\rho_{\rm
cl.}(0)-\left[\frac{3}{8\pi}h_0^2+\frac{60b'}
{7}\frac{h_0^2}{a_0^2}+6b'h_0^4+\frac{54b'}{7}\frac{1}{a_0^4}\right]\right\}.
\end{equation}

Like the quintessence case, it can be easily shown that the
resulting equations have no non-trivial solution in
$b'\rightarrow0$ limit, so there is no $\omega=-1$ crossing in
phantom model in classical level. In $b'\neq0$, $h_1$ becomes
non-trivial and is determined by eq.(\ref{28}).

In special case $\rho_m(0)=0$ and $V=0$, one has
$\rho_{\textrm{cl.}}=-(1/2)\dot\sigma^2(0)$ and therefore
eq.(\ref{28}) clearly results in:
\begin{equation}\label{29}
h_1<0,\:\:\textrm{for free pure phantom model}
\end{equation}
which proves the transition from $\omega<-1$ region to $\omega>-1$
region. In this case, the value of $h_2$ is determined by relation
\begin{eqnarray}\label{30}
h_2=\frac{1}{10368\pi b'\dot{\sigma}^4(0)a_0^4}&&\left\{
2268\pi\dot\sigma^6(0)a_0^4
+63\ddot\sigma^2(0)\dot\sigma^2(0)a_0^4 +112 \pi
b'\ddot\sigma^4(0)a_0^4 +\right. \nonumber
\\&& \left.2400\pi b'
\ddot\sigma^2(0)\dot\sigma^2(0)a_0^2 + 27216\pi b'
\dot\sigma^4(0)\right\}
\end{eqnarray}
which is a positive
quantity.

Therefore for arbitrary matter density and phantom potential, the
sign of $h_1$ and $h_2$ can be positive or negative, depending on
values $h_0$, $a_0$, $\dot\sigma(0)$, $V(0)$, and $\rho_m(0)$. But
in free ($V=0$) and pure ($\rho_m=0$) phantom model, the sign of
$h_1$ and $h_2$ are uniquely determined by Friedman equations.
$h_1$ is negative
and $h_2$ is positive. \\ \\
 {\bf Acknowledgement:} This work was partially supported by the "center of excellence in
structure of matter" of the Department of Physics of University of
Tehran, and also a
research grant from University of Tehran.\\ \\

\end{document}